\PassOptionsToPackage{colorlinks=true,allcolors=blue}{hyperref}
\documentclass[%
  aps,
  prapplied,
  reprint,
  superscriptaddress,
  longbibliography
]{revtex4-2}

\usepackage{graphicx}
\usepackage{amsmath,amssymb,bm}
\usepackage{microtype} 

\usepackage{orcidlink}
\usepackage{siunitx}
\DeclareSIUnit\sample{Sa}
\usepackage[utf8]{inputenc}
\usepackage[T1]{fontenc}
\usepackage{newunicodechar}
\newunicodechar{́}{\'}

\begin{document}
\raggedbottom

\title{Position measurement of a levitated particle with vectorial light}

\author{Daniel Tandeitnik\,\orcidlink{0000-0003-3276-9335}}
\email{tandeitnik@gmail.com}
\affiliation{Department of Physics, Pontifical Catholic University of Rio de Janeiro, Rio de Janeiro 22451-900, Brazil}

\author{Lucas Bianchi \,\orcidlink{0009-0000-2788-7578}}
\affiliation{Instituto de Física, Universidade Federal do Rio de Janeiro Caixa Postal 68528, Rio de Janeiro, Rio de Janeiro, 21941-585, Brazil}

\author{Sebastian Gutierrez-Bernal\, \orcidlink{0009-0003-7161-0498}}
\affiliation{School of Engineering and Sciences, Tecnol\'ogico de Monterrey, Monterrey, N.L., Mexico}

\author{Joanna A. Zielińska\,
\orcidlink{0000-0002-5711-1401}}
\affiliation{School of Physics and Astronomy, University of Southampton, SO17 1BJ, Southampton, UK.}

\author{Paulo A. Maia Neto}
\affiliation{Instituto de Física, Universidade Federal do Rio de Janeiro Caixa Postal 68528, Rio de Janeiro, Rio de Janeiro, 21941-585, Brazil}

\author{Thiago Guerreiro\,\orcidlink{0000-0001-5055-8481}}
\affiliation{Department of Physics, Pontifical Catholic University of Rio de Janeiro, Rio de Janeiro 22451-900, Brazil}

\date{\today}

\begin{abstract}
We develop a fully vectorial, semiclassical scattering formalism for optically levitated dipolar scatterers, expressed within the angular spectrum representation and applicable to arbitrary trapping-field configurations as well as to high--numerical-aperture focusing. Within this framework, we introduce the information radiation pattern to characterize the angular distribution of position-dependent information and use a Richards--Wolf projection of the scattered field onto the local-oscillator mode to quantify the resulting mode-matching efficiency, yielding experimentally realistic forward- and backward-detection efficiencies. As a worked example, we apply the formalism to a radially polarized trapping beam and confirm that the axial recoil heating rate is reduced relative to a conventional linearly polarized Gaussian tweezer. The theoretical framework is implemented in LevitationToolbox, an open-source Python package intended to support the design and optimization of near-Heisenberg-limited levitated optomechanical experiments.
\end{abstract}

\maketitle

\section{Introduction}

Levitated optomechanics uses optical, electric, or magnetic fields to trap a mesoscopic particle in vacuum, isolating its center-of-mass motion from a solid substrate and giving experiments access to exceptionally high mechanical quality factors~\cite{gonzalez2021levitodynamics,millen2020optomechanics}. This isolation has enabled the preparation of the motional ground state of optically trapped nanoparticles, achieved either through coherent coupling to an optical cavity~\cite{delic2020cooling,piotrowski2023simultaneous,dania2025high} or through shot-noise-limited position detection combined with active feedback~\cite{magrini2021real,tebbenjohanns2021quantum}. Ground-state control of levitated systems is now being pursued as a resource for macroscopic quantum superpositions~\cite{neumeier2024fast}, precision force and acceleration sensing~\cite{ranjit2016zeptonewton,moore2025search}, and searches for physics beyond the Standard Model~\cite{moore2021searching,afek2022coherent}.

Central to all of these applications is the optical position measurement itself. A levitated particle continuously scatters photons from the trapping field, and the interference of this scattered light with a local oscillator encodes the particle's displacement~\cite{tebbenjohanns2019optimal,tandeitnik2026heterodyne}. The same scattering process that reveals this information also imparts random momentum kicks to the particle -- measurement backaction -- so that the achievable displacement sensitivity is fundamentally limited by the interplay between imprecision and backaction noise, with their product bounded by the Heisenberg limit~\cite{tebbenjohanns2019optimal,jain2016direct}. Reaching this bound in practice requires understanding, for a given trap and collection geometry, what fraction of the position information radiated by the particle is actually captured and converted into a measurable signal.

The standard theoretical treatment of this problem models the particle as a point dipole illuminated by a single plane wave~\cite{tebbenjohanns2019optimal,seberson2020distribution}. This plane-wave approximation yields simple, widely used analytic expressions for the backaction and imprecision noise along each motional axis and has correctly predicted the strongly asymmetric, axis-dependent detection efficiencies observed in detection setups~\cite{tebbenjohanns2021quantum,magrini2021real,tandeitnik2026heterodyne}. However, real optical tweezers rely on high numerical-aperture objectives to achieve tight, stiff traps where the paraxial, scalar picture underlying the plane-wave approximation breaks down: tightly focused beams develop substantial longitudinal field components and strong depolarization at the focus~\cite{richards1959electromagnetic,wolf1959electromagnetic,novotny2012principles}, effects entirely absent from a plane-wave description. Any experiment aiming to approach the fundamental sensitivity limit therefore needs a theory that treats the vectorial, non-paraxial structure of the focal field on the same footing as the particle's dipolar response.

Two distinct routes have recently been taken to generalize the point-dipole/PWA theory. Maurer et al. developed a quantum electrodynamical theory of light-particle coupling valid for a Lorenz-Mie scatterer of arbitrary size and refractive index, going beyond the point-dipole approximation to describe recoil heating and information radiation patterns for particles ranging from the sub-wavelength to the micrometer scale~\cite{maurer2023quantum}. Their work targets the size regime relevant to scaling levitated ground-state cooling to substantially more massive particles, but retains a simple, fixed-polarization Gaussian illumination in its worked examples. The present work pursues the complementary generalization: we remain within the point-dipole (Rayleigh) regime appropriate for the sub-wavelength silica nanoparticles used in most current levitated optomechanics experiments~\cite{piotrowski2023simultaneous,dania2025high,kamba2025quantum} and instead generalize the optical field, developing a fully vectorial scattering formalism based on the angular spectrum representation that is valid for arbitrary trapping-field configurations and arbitrary numerical apertures.

This generalization is not merely formal. Because the vectorial structure of the focal field depends sensitively on its polarization and modal content, structuring the trapping beam offers a lever for engineering the spatial distribution of scattered position information -- for instance, recent work predicts that a radially polarized trapping beam can substantially suppress axial backaction relative to a conventional Gaussian tweezer~\cite{almeida2025levitated}. Testing and exploiting such predictions requires a framework that (i) self-consistently accounts for high-NA focusing and depolarization, (ii) applies to arbitrary superpositions of vector beam modes rather than a single beam profile, and (iii) connects the resulting information radiation pattern to what a realistic phase-sensitive detector -- with a finite collection aperture and an imperfectly mode-matched local oscillator -- can actually measure.

In this article, we develop such a framework. We derive a coherent scattering vector that captures the interference between the incident momentum distribution and the dipole emission pattern for an arbitrary non-paraxial trapping field and use it to define a generalized geometric factor that fixes both the backaction and imprecision noise while preserving the Heisenberg limit. We introduce the information radiation pattern as the angular density of position information radiated by the particle and, by projecting the scattered field onto the local-oscillator mode via the Richards--Wolf formalism, quantify the resulting mode-matching efficiency to obtain, beyond simple geometric collection efficiency, the hardware detection efficiency achievable in realistic forward- and backward-detection schemes. We benchmark the formalism against the known plane-wave approximation point-dipole results and apply it to compare a standard linearly polarized Gaussian tweezer against a radially polarized trap, showing how focal depolarization and field structuring reshape both the backaction and the achievable measurement efficiency.

\section{Theoretical Framework: The Semi-Classical Model}

\subsection{System Dynamics}

In the semiclassical model developed in this section, we introduce two primary assumptions. First, we assume that the radius $a$ of the scatterer is much smaller than the wavelength $\lambda$ of the trapping field ($a \ll \lambda$), placing the system in the Rayleigh scattering regime. Under this condition, the particle's electromagnetic response is dominated by the lowest-order electric dipole term of the Lorenz--Mie expansion, with higher-order multipoles suppressed by successive powers of $a/\lambda$~\cite{maurer2023quantum}. By specifically considering a spherical particle composed of an isotropic material, such as amorphous silica, the scatterer can be treated as an effective point dipole characterized solely by a scalar polarizability $\alpha$, avoiding the need for a full polarizability tensor. Second, the term semiclassical here indicates that the trapping field is treated as being in a coherent state, such that the photon scattering events by the particle follow Poissonian statistics.

Consider an optically levitated particle of mass \(m\), confined within a harmonic potential generated by the trapping field. Its dynamics along a generic spatial coordinate \(q \in \{x,y,z\}\) are described by the Langevin equation
\begin{equation}
    \ddot{q}(t) + \gamma_{\mathrm{th}} \dot{q}(t) + \Omega_q^2 q(t) = \frac{1}{m}\big[F_{\mathrm{th},q}(t) + F_{\mathrm{bk},q}(t)\big],
\end{equation}
where \(\Omega_q\) denotes the mechanical resonance frequency and \(\gamma_{\mathrm{th}}\) is the damping rate arising from residual gas collisions. Laser-induced damping, \(\gamma_{\mathrm{bk}} = 2P/(mc^2)\)~\cite{karrai2008doppler,jain2016direct}, where $P$ is the trapping field power, is neglected since it is dominated by gas damping at medium vacuum pressures and by active feedback control in the ultra-high-vacuum regime. The thermal Brownian force \(F_{\mathrm{th},q}(t)\) is modeled as a Gaussian white noise process with a (one-sided) power spectral density (PSD) given by \(S_{\mathrm{th},q} = 4 m \gamma_{\mathrm{th}} k_B T\), where \(k_B\) is the Boltzmann constant and \(T\) is the bath temperature.
\begin{figure}
    \centering
    \includegraphics{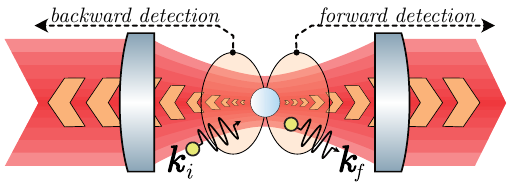}
    \caption{Schematic representation of coherent photon scattering and position detection in an optical levitation setup. The trapping beam (propagating left to right) is strongly focused by a high-NA objective. The levitated nanoparticle acts as a driven point-dipole, interacting with an incident photon (wavevector $\mathbf{k}_i$) and scattering it into a final state (wavevector $\mathbf{k}_f$). The resulting dipole radiation pattern carries the particle's spatial information into the far-field. This coherent signal is subsequently gathered either by the trapping objective (backward detection) or by a secondary collection lens (forward detection) for phase-sensitive measurement.}
    \label{fig:scatteringModel}
\end{figure}

The measurement backaction force $F_{\mathrm{bk},q}(t)$ is also modeled as a white Gaussian noise process $F_{\mathrm{bk},q}(t) = \sigma_{\mathrm{bk},q}\eta_{\mathrm{bk},q}(t)$, where $\eta_{\mathrm{bk},q}(t)$ is a zero-mean, unit-variance white noise process satisfying $\langle\eta_{\mathrm{bk},q}(t)\eta_{\mathrm{bk},q}(t+\tau)\rangle = \delta(\tau)$, and $\sigma_{\mathrm{bk},q}$ is the corresponding noise amplitude, to be determined below in terms of the photon scattering statistics. While the spatial distribution of scattered light is governed by coherent wave interference, the backaction force itself is fundamentally stochastic due to the quantized nature of the electromagnetic field. Because the trapping laser is a coherent state, discrete photon scattering events follow independent Poissonian statistics (photon shot noise)~\cite{gerry2023introductory}.

By the Wiener–Khinchin theorem, the backaction noise is frequency-independent and therefore given by \(S_{\mathrm{bk},q} = \sigma_{\mathrm{bk},q}^2\). This variance is directly related to the momentum diffusion rate via~\cite{clerk2010introduction}
\begin{equation}\label{eq:PSDdiffRel}
    S_{\mathrm{bk},q} = \frac{\Delta p_{\mathrm{bk},q}^2}{\tau},
\end{equation}
where \(\Delta p_{\mathrm{bk},q}^2\) denotes the momentum variance accumulated over an integration time $\tau$.

\subsection{Coherent Scattering and Momentum Transfer}

Modeling photon scattering as a compound stochastic process, the total momentum transferred along a given axis over a time interval \(\tau\) is
\begin{equation}
    P_{\text{tot}} = \sum_{j=1}^{N} p_j,
\end{equation}
where \(p_j\) is the momentum imparted by the \(j\)-th photon, and \(N\) is the random total number of scattered photons. The variables \(p_j\) and \(N\) are taken to be statistically independent. For a coherent laser source, the photon-number statistics are Poissonian, such that \(\mathrm{Var}(N) = \langle N \rangle\)~\cite{gerry2023introductory}. Applying the law of total variance~\cite{papoulis1967probability} to \(P_{\text{tot}}\) yields
\begin{align}
    \mathrm{Var}(P_{\text{tot}}) &= \mathbb{E}\big[\mathrm{Var}(P_{\text{tot}} \mid N)\big] 
    + \mathrm{Var}\big(\mathbb{E}[P_{\text{tot}} \mid N]\big) \nonumber \\
    &= \langle N \rangle \big(\langle p^2 \rangle - \langle p \rangle^2\big) 
    + \langle p \rangle^2 \,\mathrm{Var}(N).
\end{align}
Substituting the Poissonian relation between the variance and the mean of $N$, the contributions involving the mean momentum transfer cancel exactly, resulting in
\begin{equation}\label{eq:totalVariance}
    \mathrm{Var}(P_{\text{tot}}) = \langle N \rangle \langle p^2 \rangle.
\end{equation}
Thus, the total momentum variance is strictly proportional to the second moment of the individual momentum kicks in the coherent field.

In the Rayleigh scattering regime, the nanoparticle can be modeled as an effective point electric dipole driven by the coherently superposed focal field $\mathbf{E}_{\mathrm{inc}}$. The resulting local driving field is given by
\begin{equation}
    \mathbf{E}_{\mathrm{local}} = \int \mathbf{E}_{\mathrm{inc}}(\mathbf{k}_i)\, d\Omega_i,
\end{equation}
where the integral is taken over the solid angle of incident wave vectors $\mathbf{k}_i$. When a photon is scattered into a final direction $\mathbf{k}_f$, the associated momentum transfer amplitude is obtained by integrating the difference of the wave vectors, weighted by the complex amplitudes of the incident field (see Fig.~\ref{fig:scatteringModel}). We define this quantity as the coherent scattering vector $\mathbf{V}_q$:
\begin{align}
    \mathbf{V}_q(\mathbf{k}_f) &= \int (k_{i,q} - k_{f,q})\, \mathbf{E}_{\mathrm{inc}}(\mathbf{k}_i)\, d\Omega_i \nonumber\\&\equiv \mathbf{W}_q - k_{f,q}\mathbf{E}_{\mathrm{local}},
\end{align}
where $\mathbf{W}_q$ represents the contribution associated with the incident momentum distribution. This expression naturally decomposes into an effective extinction term $\mathbf{W}_q$ proportional to the spatial gradient of the focal field, and a scattered momentum term $k_{f,q}\mathbf{E}_{\mathrm{local}}$.

Physically, $\mathbf{W}_q$ represents the spatial derivative of the local electric field. This derivative naturally splits into amplitude and phase gradients, driving two distinct momentum transfer mechanisms. The amplitude gradient couples with the particle's real polarizability to form the conservative trapping potential. The phase gradient, conversely, defines the local optical momentum flux. Even when intrinsic material absorption is negligible, the nanoparticle extracts momentum from this flux via elastic scattering. To conserve energy, this scattering induces an effective imaginary radiation-reaction polarizability~\cite{jones2015optical}. This imaginary component couples to the phase gradient to mediate all non-conservative forces, ultimately dictating both the mean radiation pressure and the random momentum fluctuations responsible for measurement backaction.

To enforce the transversality condition of far-field electromagnetic radiation~\cite{griffiths2023introduction}, we project the coherent scattering vector onto the plane orthogonal to $\mathbf{k}_f$:
\begin{equation}
    \mathbf{V}_q^\perp = \mathbf{W}_q^\perp - k_{f,q}\mathbf{E}_{\mathrm{local}}^\perp,
\end{equation}
where the superscript $\perp$ denotes the transverse component with respect to $\mathbf{k}_f$. The dipole radiation pattern emerges directly from this three-dimensional vector construction. Specifically, by taking the squared modulus of the transverse emission term, factoring out the local field magnitude, and introducing the unit polarization vector $\hat{\mathbf{p}}$, one obtains $|\mathbf{E}_{\mathrm{local}}|^2 \bigl(1 - |\hat{\mathbf{p}}\cdot \hat{\mathbf{k}}_f|^2\bigr)$, which corresponds to the standard angular dependence of dipole radiation~\cite{jackson2012classical}. By evaluating the full squared magnitude $|\mathbf{V}_q^\perp|^2$, we obtain a description of the scattered intensity that consistently incorporates the dipole emission pattern, the distribution of incident momenta, and their coherent interference, while strictly satisfying the geometric and polarization constraints of far-field radiation.

\subsection{Angular Spectrum Representation}

To accurately evaluate the coherent scattering vector $\mathbf{V}_q$ in the non-paraxial regime, it is necessary to go beyond scalar approximations. In conventional levitated optomechanics, particles are confined using high–numerical-aperture (NA) objectives, for which the paraxial approximation is no longer valid. Consequently, we construct the incident field using the fully vectorial Angular Spectrum Representation (ASR)~\cite{wolf1959electromagnetic,richards1959electromagnetic,novotny2012principles}. Furthermore, the ASR formalism naturally accommodates arbitrary trapping fields, enabling the direct analysis of complex vectorial optical potentials and alternative trapping geometries.

\begin{figure}
    \centering
    \includegraphics{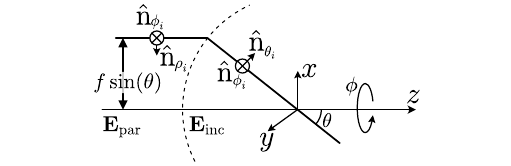}
    \caption{Geometrical representation of the coordinate transformations at the reference sphere of an aplanatic lens with focal length $f$. The incident paraxial field $\mathbf{E}_{\mathrm{par}}$ propagates along the $z$-axis. Upon refraction, the radial basis vector $\hat{\mathbf{n}}_{\rho_i}$ rotates to become the polar unit vector $\hat{\mathbf{n}}_{\theta_i}$ for the corresponding Fourier component of the focused field $\mathbf{E}_{\mathrm{inc}}$. The azimuthal vector $\hat{\mathbf{n}}_{\phi_i}$ (pointing into the page) remains orthogonal to the plane of incidence and is invariant under focusing.}
    \label{fig:refractionRefplane}
\end{figure}

We define the angular spectrum $\mathbf{E}_{\mathrm{inc}}(\theta_i, \phi_i)$ by mapping the incident paraxial transverse field $\mathbf{E}_{\mathrm{par}}$ through an aplanatic lens of focal length $f$. Such a lens obeys the sine condition~\cite{novotny2012principles}, which maps the radial coordinate $\rho$ of the incident paraxial beam to the convergence angle $\theta_i$ via the relation $\rho = f \sin\theta_i$. Upon refraction, the azimuthal basis vector $\hat{\mathbf{n}}_{\phi_i}$ remains invariant, as it is orthogonal to the plane of incidence, whereas the radial basis vector $\hat{\mathbf{n}}_{\rho_i}$ is rotated into the polar unit vector $\hat{\mathbf{n}}_{\theta_i}$, which is tangent to a reference sphere centered at the lens focus; see Fig.~\ref{fig:refractionRefplane} for a schematic representation. The resulting field on the reference sphere is given by
\begin{align}
    \mathbf{E}_{\mathrm{inc}}(\theta_i&,\phi_i) =\nonumber\\
    &\sqrt{\cos\theta_i} \left[ (\mathbf{E}_{\mathrm{par}}\cdot\hat{\mathbf{n}}_{\phi_i})\hat{\mathbf{n}}_{\phi_i} + (\mathbf{E}_{\mathrm{par}}\cdot\hat{\mathbf{n}}_{\rho_i})\hat{\mathbf{n}}_{\theta_i} \right],
\end{align}
\noindent where, for clarity, we assume that the refractive index is identical on both sides of the lens, and the prefactor $\sqrt{\cos\theta_i}$ arises from enforcing energy conservation in the focusing process~\cite{novotny2012principles}.

Moreover, within the ASR formalism, it is necessary to account for the finite numerical aperture of the lens, which determines the maximum acceptance angle $\theta_{\mathrm{max}}$ for which an incoming plane wave can be refracted to the focal point, according to the relation $\mathrm{NA} = \sin(\theta_{\mathrm{max}})$. For a given incident field with waist $w_0$, we define the filling factor $f_0$ as
\begin{equation}
    f_0 = \frac{w_0}{f\sin(\theta_{\mathrm{max}})} = \frac{w_0}{f\,\mathrm{NA}},
\end{equation}
\noindent The filling factor is a key parameter characterizing the incident beam relative to the lens pupil. It determines the fraction of power that is clipped at the lens entrance, and the transverse as well as longitudinal structure of the focused field can vary substantially as a function of $f_0$ for structured fields.

\subsection{Generalized geometric factor}

To obtain a purely geometric and dimensionless measure of the variance associated with the coherent scattering vector, we integrate its transverse intensity over all final solid angles $\Omega_f$ and apply three successive normalizations. First, we divide by the local field intensity $|\mathbf{E}_{\mathrm{local}}|^2$, thereby removing the dependence on the absolute optical power and isolating the geometrical structure of the focal field. Second, we divide by the squared wavenumber $k^2$ to express the momentum spread as a dimensionless fraction of a single photon's total momentum. Third, we convert the resulting integrated intensity into a properly normalized probability density function. 

Within classical electrodynamics, the spatial integral of a normalized dipole radiation pattern over the full solid angle of a sphere yields a total value of $8\pi/3$~\cite{jackson2012classical}. Consequently, dividing by this emission volume imposes the correct probabilistic normalization on the statistics. Under these conditions, we define the generalized geometric factors $\mathrm{C}_q$ as
\begin{equation}\label{eq:genGeoFactor}
    \mathrm{C}_q = \frac{3}{8\pi} \frac{1}{k^2 |\mathbf{E}_{\mathrm{local}}|^2} \int_{\Omega_f} \bigl|\mathbf{V}_q^\perp(\mathbf{k}_f)\bigr|^2 \, d\Omega_f.
\end{equation}

The generalized geometric factor fully characterizes the momentum transfer between the trapping field and the scatterer in the non-paraxial regime for an arbitrary trapping configuration. It inherently accounts for the incident momentum distribution, the dipolar emission pattern specified by $\mathbf{E}_{\mathrm{local}}$, and the mutual interference between these contributions.
\begin{figure*}
    \centering
    \includegraphics{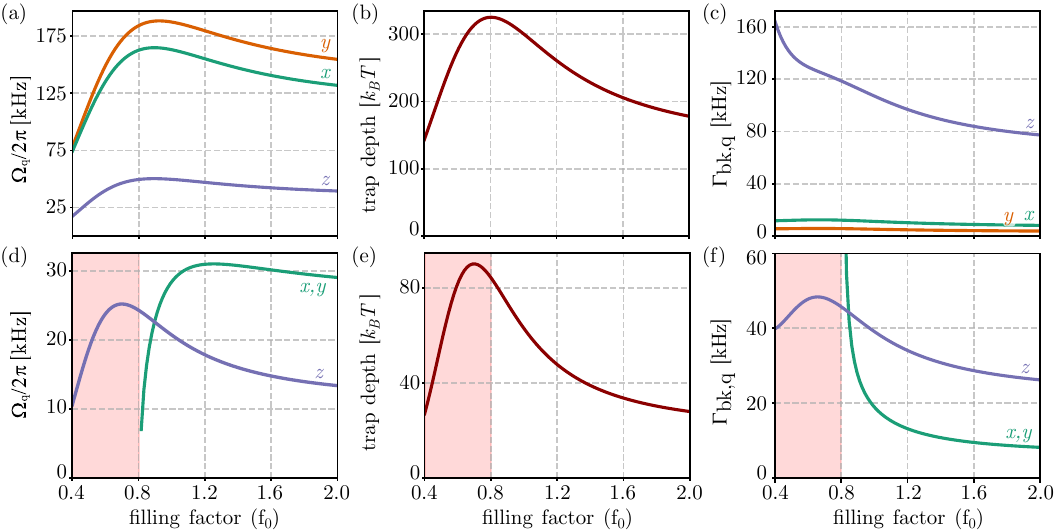}
    \caption{Comparison of optomechanical trapping parameters for a linearly polarized fundamental Gaussian beam (top row, a–c) and a radially polarized beam (bottom row, d–f) as a function of the objective filling factor $f_0$. All calculations assume a total incident laser power of $\SI{1}{\watt}$ prior to the trapping objective ($\text{NA} = 0.75$), and a $\mathrm{SiO}_2$ particle with diameter $\SI{156}{\nano\meter}$. (a, d) Mechanical trap frequencies $\Omega_q / 2\pi$. (b, e) Trap depth at the focal origin, normalized by the thermal energy $k_B T$ at $\SI{293}{\kelvin}$. (c, f) Backaction decoherence rates $\Gamma_{\mathrm{bk},q}$. The light red shaded regions denote unstable parameter regimes. For the radial mode, this instability occurs at lower filling factors ($f_0 \lesssim 0.8$) where the absence of highly convergent marginal rays prevents the formation of a central intensity maximum.}
    \label{fig:gaussRadPanelDeco}
\end{figure*}
\section{Fundamental noise limits}

\subsection{Backaction and imprecision noise}

By expressing the total momentum variance in terms of $\mathrm{C}_q$, we obtain
\begin{equation}
    \Delta p_{\mathrm{bk},q}^2 = \langle N \rangle \hbar^2 k^2 \mathrm{C}_q,
\end{equation}
\noindent where Eq.~\eqref{eq:totalVariance} has been used to account for $\langle N \rangle$ independent scattering events. Introducing the photon scattering rate $\Gamma_{\mathrm{S}} = \langle N \rangle / \tau$ and substituting this expression for the variance into Eq.~\eqref{eq:PSDdiffRel} yields an explicit form for the backaction force spectral density:
\begin{equation}
    S_{\mathrm{bk},q} = \mathrm{C}_q \hbar^2 k^2 \Gamma_{\mathrm{S}}.
\end{equation}

A displacement of the particle by $\Delta q$ induces a relative phase shift in the scattered field that is proportional to the momentum transfer amplitude~\cite{tebbenjohanns2019optimal}. The total phase variance $\Delta \Phi_q^2$ accessible in the far-field assumes a mathematical form that is exactly analogous to the momentum variance:
\begin{align}
    \Delta \Phi_q^2 &= \Delta q^2 \left( \frac{3}{8\pi} \frac{1}{\vert{}\mathbf{E}_{\mathrm{local}}\vert{}^2} \int_{\Omega_f} \vert{}\mathbf{V}_q^\perp(\mathbf{k}_f)\vert{}^2 \, d\Omega_f \right) \nonumber\\&= \mathrm{C}_q k^2 \Delta q^2.
\end{align}
\noindent Comparing this signal variance to the intrinsic phase shot noise of a coherent probe field, $\Delta \Phi^2 = 1/(4\langle N\rangle)$~\cite{gerry2023introductory}, yields the corresponding measurement imprecision noise:
\begin{equation}
    S_{I,q} = \left( 4 \mathrm{C}_q k^2 \Gamma_{\mathrm{S}} \right)^{-1}.
\end{equation}
\noindent The product of the measurement imprecision and the associated backaction noise then yields $S_{I,q} S_{\mathrm{bk},q} = \hbar^2/4$, showing that the Heisenberg limit is exactly saturated, irrespective of the spatial mode structure of the probe beam or the degree of non-paraxial focusing.

We therefore identify the generalized geometric factors $\mathrm{C}_q$ as key quantities that govern the characterization of noise processes within the system. Under the plane wave approximation (PWA), i.e., when the incident field is modeled as a single, spatially uniform plane wave propagating along the optical axis $\mathbf{\hat{z}}$ and, for instance, linearly polarized along $\mathbf{\hat{x}}$, we recover the well-known result of Refs.~\cite{tebbenjohanns2019optimal,seberson2020distribution}, namely $\mathbf{C} = \left[0.2,\; 0.4,\; A^2 + 0.4\right]$, via Eq.~\eqref{eq:genGeoFactor}. Here, $A$ denotes a correction factor that accounts for the effect of the Gouy phase identified in Ref.~\cite{tebbenjohanns2019optimal}. For a general incident field, the generalized definition of $\mathrm{C}_q$ naturally incorporates and accurately reflects the modified optical configuration.

\subsection{Backaction decoherence rate}

The measurement backaction fundamentally adds phonons to the oscillator. The backaction decoherence rate, also termed recoil heating, is found by multiplying the mean energy transfer per photon by the scattering rate~\cite{jain2016direct,seberson2020distribution}:
\begin{equation}\label{eq:decoherenceRate}
    \Gamma_{\mathrm{bk},q} = \mathrm{C}_q \frac{\hbar^2k^2}{2m\hbar\Omega_q}\Gamma_{\mathrm{S}}.
\end{equation}

To demonstrate the applicability of the generalized ASR framework, we evaluate and compare the optomechanical parameters associated with two distinct optical trapping configurations: a standard linearly polarized fundamental Gaussian beam and a radially polarized beam. The latter, characterized by a toroidal intensity profile in the paraxial limit, can be represented as a superposition of two orthogonal first-order Hermite–Gaussian ($\mathrm{HG}_{mn}$) modes:
\begin{equation}
    \mathbf{E}_{\mathrm{rad}} 
    = \mathrm{HG}_{10}\,\hat{\mathbf{x}} + \mathrm{HG}_{01}\,\hat{\mathbf{y}}.
\end{equation}
\noindent The former corresponds to the conventional configuration employed in optical tweezers, whereas the latter constitutes a highly structured optical vortex field. The radial beam was numerically predicted in Ref.~\cite{almeida2025levitated} to exhibit a reduced decoherence rate along the axial direction relative to the former. This prediction was obtained using a MATLAB toolbox~\cite{callegari2015computational} to simulate the interaction between the trapping field and the particle in the Mie-scattering regime. Furthermore, the ASR formalism becomes essential for the analysis of the radially polarized beam, as it predicts that, under high-NA focusing, its donut-shaped intensity distribution collapses into a bright on-axis spot that is predominantly polarized along the axial direction, thereby generating a confining optical potential.

Fig.~\ref{fig:gaussRadPanelDeco} presents the mechanical trap frequencies, trap depths, and backaction-induced decoherence rates as functions of the objective filling factor $f_0$. The upper row corresponds to a linearly polarized Gaussian trapping beam, whereas the lower row corresponds to a radially polarized trapping beam. Both configurations are investigated for a trapping field with wavelength $\SI{1550}{\nano\meter}$ and incident laser power $\SI{1}{\watt}$ specified at the entrance pupil of an objective with numerical aperture $\mathrm{NA} = 0.75$~\footnote{The quantities shown scale proportionally to the square root of the incident power. This scaling can therefore be used to rescale the results to any other specified power.}. The trapped object is a $\mathrm{SiO_2}$ nanoparticle with diameter $\SI{156}{\nano\meter}$.

To evaluate these observables, we first compute the three-dimensional distribution of the local electric-field energy density, $|\mathbf{E}(\mathbf{r})|^2$, by propagating the incident angular spectrum through the aplanatic objective. The time-averaged optical dipole potential is then given by
\begin{equation}
    U(\mathbf{r}) = -\frac{1}{4}\,\alpha\,|\mathbf{E}(\mathbf{r})|^2,
\end{equation}
where $\alpha$ denotes the particle polarizability, which is expressed via the Clausius--Mossotti relation~\cite{jones2015optical} as
\begin{equation}
    \alpha = 3V\epsilon_0\frac{\epsilon_r-1}{\epsilon_r+2},
\end{equation}
with $V$ the particle volume, $\epsilon_0$ the vacuum permittivity, and $\epsilon_r \equiv \epsilon_p / \epsilon_m$ the relative permittivity, defined as the ratio of the particle permittivity $\epsilon_p$ to that of the surrounding medium $\epsilon_m$. The particle is trapped at the equilibrium position $\mathbf{r}_{\mathrm{eq}}$, defined as a local minimum of $U(\mathbf{r})$. Provided $\mathbf{r}_{\mathrm{eq}}$ is a global minimum of $U(\mathbf{r})$, the trap depth is defined as $|U(\mathbf{r}_{\mathrm{eq}})|$. For all configurations considered in this work, we have verified numerically that this condition holds along all three axes. The mechanical resonance frequencies are similarly obtained from the harmonic trap stiffness, determined by the spatial curvature of $|\mathbf{E}(\mathbf{r})|^2$ at $\mathbf{r}_{\mathrm{eq}}$. Since the particle is a non-absorptive dielectric, the photon-scattering rate at equilibrium is calculated as
\begin{equation}
    \Gamma_{\mathrm{S}} = \frac{\sigma_{\mathrm{ext}} |\mathbf{E}(\mathbf{r}_{\mathrm{eq}})|^2}{\hbar k c},
\end{equation}
where $\sigma_{\mathrm{ext}}$ is the extinction cross-section, well approximated in this regime by the Rayleigh-scattering expression~\cite{jones2015optical}:
\begin{equation}
    \sigma_{\mathrm{ext}} \approx \frac{8\pi^3\alpha^2}{3\epsilon_0^2\lambda^4}.
\end{equation}

Owing to the intrinsic cylindrical symmetry of the radially polarized mode, the transverse mechanical eigenfrequencies are degenerate, in contrast to the symmetry breaking observed in a linearly polarized Gaussian trap~\cite{novotny2012principles}. Although the radially polarized beam produces a smaller overall trapping potential depth and reduced mechanical frequencies for the same incident optical power, it exhibits a pronounced advantage with respect to measurement-induced backaction. 

In particular, the axial decoherence rate $\Gamma_{\mathrm{bk},z}$ is significantly reduced relative to the Gaussian configuration. In the worked example, for the case of $f_0 = 1$, the model predicts a reduction by a factor of approximately $2.7$. Within the ASR framework, this suppression is quantitatively captured by the angular integration of the coherent scattering vector $\mathbf{V}_z^\perp$. When a radially polarized beam is tightly focused by a high-numerical-aperture objective, constructive interference gives rise to a purely longitudinal local electric field at the trap center. The emission contribution to the axial scattering vector is determined by the term $k_{f,z}\mathbf{E}_{\mathrm{local}}^\perp$. Because the nanoparticle behaves as a dipole oriented along the $z$-axis, it radiates predominantly into the transverse plane, where the longitudinal momentum component of the scattered photons approaches zero. As a result, the maximum of the scattered intensity coincides precisely with the node of the axial momentum transfer. This geometric orthogonality fundamentally minimizes the variance integral that defines $C_z$, providing the mechanism underlying the strongly suppressed axial decoherence rate.
\begin{figure*}
    \centering
    \includegraphics{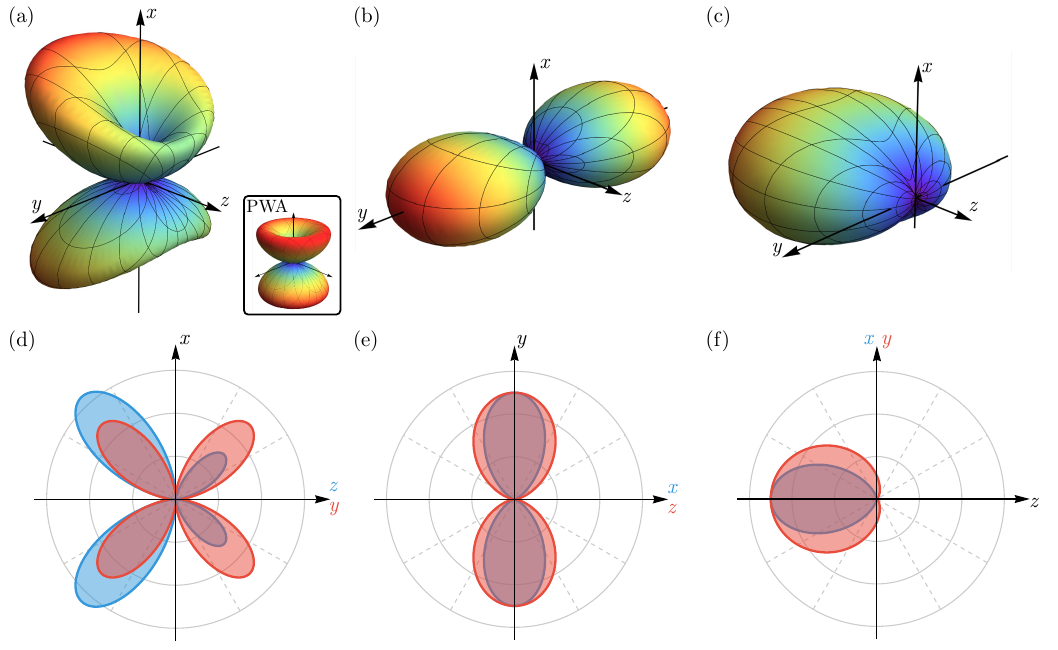}
    \caption{Spatial distribution of the position information radiated by a dipolar scatterer trapped in a strongly focused fundamental Gaussian beam (linearly polarized along the $x$-axis). (a)-(c) Three-dimensional IRP for displacements along the $x$, $y$, and $z$ axes. (d)-(f) Two-dimensional polar cross-sections of the 3D patterns, evaluated in the $xz$-plane (blue) and $xy$-plane or $yz$-plane (red). Inset in (a) shows the PWA prediction.}
    \label{fig:infoGaussian}
\end{figure*}

\section{Measurement efficiency in realistic systems}

To evaluate how a realistic experimental setup departs from the absolute Heisenberg limit, we evaluate the information gathering limit imposed by the finite numerical aperture of the collection lens, alongside detector mode mismatch and downstream optical loss.

\subsection{Information radiation pattern}
\begin{figure*}
    \centering
    \includegraphics{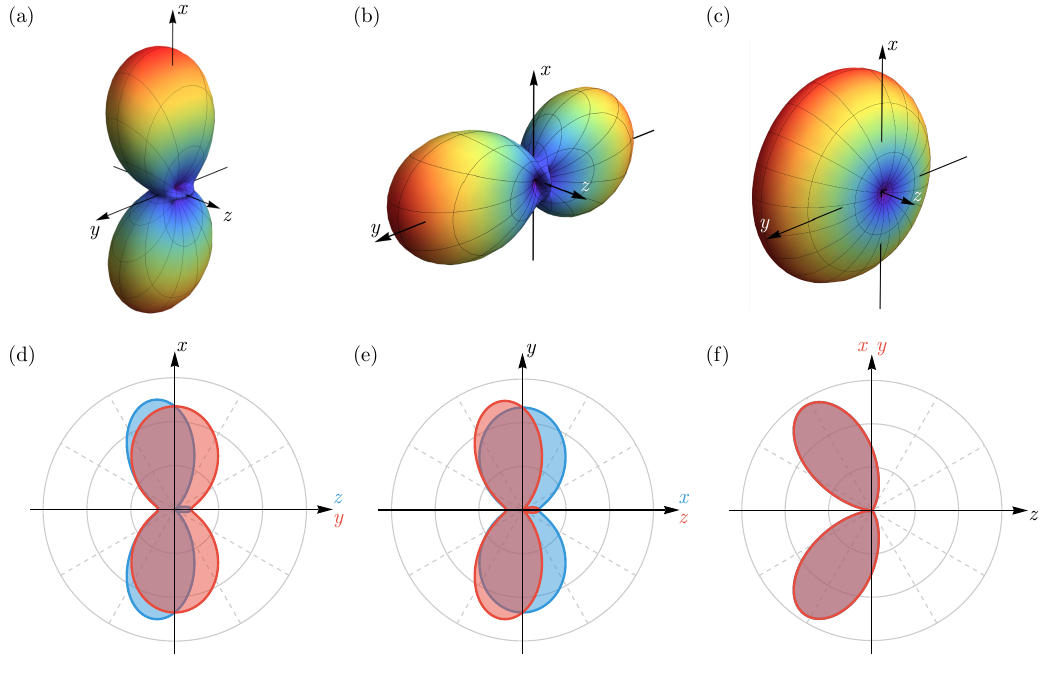}
    \caption{Spatial distribution of the position information radiated by a dipolar scatterer trapped in a strongly focused radial beam. (a)-(c) Three-dimensional Information Radiation Patterns for the $x$, $y$, and $z$ axes. (d)-(f) Two-dimensional polar cross-sections.}
    \label{fig:infoRadial}
\end{figure*}

Not all scattered photons contribute equally to position readout; depending on their scattering angle, some encode substantial phase information about the particle’s position, whereas others carry negligible positional information~\cite{tebbenjohanns2019optimal}. In a typical experimental configuration, the scattered photons are collected by a collection lens and directed to a phase-readout detection scheme~\cite{magrini2021real,moore2025search,tandeitnik2026heterodyne}. Commonly, the collection lens is either the focusing lens that generates the trapping field or a separate collimating lens placed downstream of the trap. The former collects the backward-scattered field, while the latter collects the forward-scattered field. See Fig.~\ref{fig:scatteringModel}. For either geometry, we quantify the amount of positional information captured by the lens by evaluating the coherent phase integral restricted to the solid angle subtended by the collection lens $\Omega_{\mathrm{cl}}$:
\begin{align}
    \Delta \Phi_q^{2\,\mathrm{eff}} &= \Delta q^2 \left( \frac{3}{8\pi} \frac{1}{ |\mathbf{E}_{\mathrm{local}}|^2} \int_{\Omega_{\mathrm{cl}}} |\mathbf{V}_q^\perp(\mathbf{k}_f)|^2 \, d\Omega_f \right) \nonumber\\&= \eta_{\mathrm{info},q}\,\Delta \Phi_q^{2}.
\end{align}
\noindent Here, $\eta_{\mathrm{info},q}$ denotes the information-gathering efficiency associated with motional axis $q$. It quantifies the absolute fraction of the Heisenberg-limited positional information that is physically transmitted into the acceptance cone of the collection lens.
\begin{figure*}
    \centering
    \includegraphics{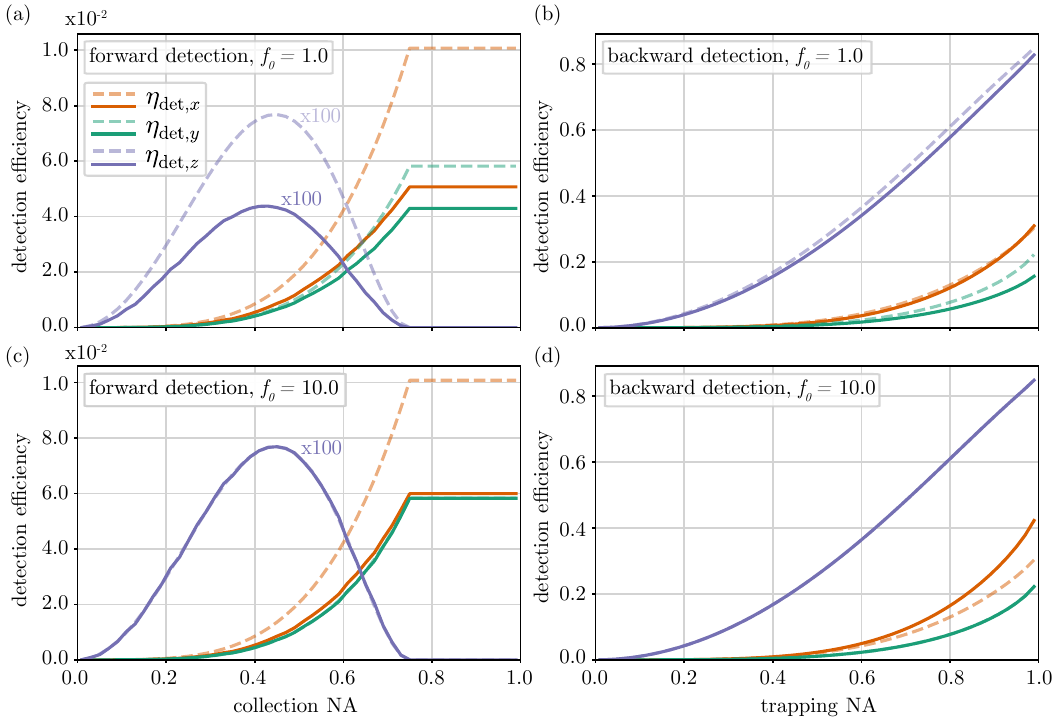}
    \caption{Comparison of detection efficiencies $\eta_{\mathrm{det},q} $ evaluated using the ASR (solid lines) and analytical PWA (dashed lines) for a linearly ($x$-)polarized trapping beam. (a, b) Forward and backward detection efficiencies for a Gaussian beam with filling factor $f_0 = 1.0$. (c, d) Efficiencies in the uniform aperture limit ($f_0 = 10.0$). In the forward detection panels (a, c), the trapping NA is fixed at 0.75, yielding peak efficiencies once the collection NA captures the entire scattered cone. Note that $\eta_z$ is scaled by a factor of $100$ for visibility in the forward direction.}
    \label{fig:measurementEfficiency}
\end{figure*}

Inspection of this integral motivates the definition of a differential information density. We define the fractional amount of information regarding the particle's displacement along axis $q$ contained within a specific solid angle $d\Omega_f$ as the information radiation pattern (IRP)~\cite{tebbenjohanns2019optimal,maurer2023quantum}:
\begin{equation}
    d\eta_{\mathrm{info},q}(\mathbf{k}_f) = \frac{|\mathbf{V}_q^\perp(\mathbf{k}_f)|^2}{\int_{4\pi} |\mathbf{V}_q^\perp(\mathbf{k}_f')|^2 d\Omega_f'} d\Omega_f.
\end{equation}
\noindent The square magnitude of the coherent scattering vector $|\mathbf{V}_q^\perp|^2$ captures the interference between the coherent local field driving the dipole and the spatial momentum gradients of the incident beam, which dictate the spatial asymmetry of the IRP. Physically, the fractional information radiated in a given direction is strictly proportional to the net momentum extracted by the photon along the measurement axis $q$. If the incident optical field possesses a strong momentum gradient along $q$, the scattered photon must carry away a commensurate momentum kick to register a measurable phase shift. Consequently, scattering directions where the photon's emission momentum constructively adds to the field's local momentum gradient will radiate the highest density of positional information, whereas directions where these momenta destructively cancel will carry negligible information. 

The spatial distribution of information can be visualized by generating surface plots of the radial density $d\eta_{\mathrm{info},q}/d\Omega_f$~\cite{tebbenjohanns2019optimal,maurer2023quantum}. Analyzing these distributions within the fully vectorial ASR framework reveals critical differences between measurement axes that are not captured by simplified scalar models.

Figure~\ref{fig:infoGaussian} presents a panel depicting the IRP for a tightly focused Gaussian beam with a numerical aperture $\mathrm{NA} = 0.75$ and filling factor $f_0 = 1$, linearly polarized along $\mathbf{\hat{x}}$. For displacements along the axis defined by the trapping polarization, the idealized plane-wave approximation predicts the familiar, symmetric figure-eight dipole radiation pattern (inset of Fig.~\ref{fig:infoGaussian}a). In contrast, the exact ASR calculation (Fig.~\ref{fig:infoGaussian}a) exhibits a pronounced narrowing of the primary lobes along the optical $z$ axis. This effect constitutes a direct visual signature of focal depolarization. 

In a high-NA focusing geometry, the strongly convergent marginal rays induce a rotation of the incident $x$-polarized field, thereby generating a substantial longitudinal, $z$-polarized component at the focus. This reorientation diverts a portion of the scattering amplitude into orthogonally polarized channels, which in turn diminishes the visibility of forward–backward interference and transforms the principal dipolar lobe into a four-lobed structure.

Conversely, for axial displacements, the spatial distribution is governed by the exceptionally large momentum gradient $\mathbf{W}_z$ established by the beam's propagation along the optical axis. The cross-term between $\mathbf{W}_z$ and $\mathbf{k}_{f,z}\mathbf{E}_{\mathrm{local}}$ generates a massive spatial asymmetry. During forward scattering ($k_{f,z} > 0$), these momentum vectors share the same sign and destructively interfere, effectively suppressing the positional information. During backward scattering ($k_{f,z} < 0$), however, the emission momentum vector flips, resulting in highly constructive interference. This massive coherent momentum recoil maximizes the displacement information carried by the back-scattered field, visually manifesting as a singular, dominant lobe pointing opposite to the direction of beam propagation (Fig.~\ref{fig:infoGaussian}f).

Because the interference is dictated by the vector geometry of the focus, the spatial distribution of information can be radically altered by structuring the incoming optical field. Figure~\ref{fig:infoRadial} presents the IRP for a particle trapped in a tightly focused radial beam. Due to the perfect rotational symmetry of the radial polarization, the transverse electric field perfectly cancels at the origin, yielding a local driving field $\mathbf{E}_{\mathrm{local}}$ that is strictly $z$-polarized. For transverse displacements, the momentum gradients are strictly orthogonal to this driving field resulting in the perfectly symmetric, four-leaf clover geometries seen in Fig.~\ref{fig:infoRadial}(a,d) and (b,e). For longitudinal motion, the back-scattering interference remains highly constructive, but the angular distribution of the incident rays hollows out the center of the lobe, creating a distinct back-directed toroidal geometry (Fig.~\ref{fig:infoRadial}(c,f)).

\subsection{Detection efficiency}

In practical experimental realizations, the scattered field is typically measured using spatially integrating detection schemes, such as balanced homodyne or heterodyne detection~\cite{tandeitnik2026heterodyne}. In these configurations, the achievable measurement efficiency is fundamentally constrained by spatial and polarization mode mismatches between the scattered radiation and the reference local oscillator (LO). A description of this overlap requires explicit consideration of the collection optics, in particular the action of the collection lens, which transforms the far-field spherical wavefront into a collimated transverse beam. 

The phase-readout signal $I_q$ arises solely from the interference cross-term between this collimated scattered field and the LO field and is further weighted by the spatial split-mask $\mathcal{M}_q(\mathbf{k}_f)$ of the detector:
\begin{equation}
    I_q \propto \frac{1}{\lvert \mathbf{E}_{\mathrm{local}} \rvert} \int_{\Omega_{\mathrm{cl}}} \mathrm{Re}\left\{ \mathbf{E}_{\mathrm{LO}}^* \cdot \mathcal{T}\big[\mathbf{V}_q^\perp(\mathbf{k}_f)\big] \right\} \mathcal{M}_q(\mathbf{k}_f)\, d\Omega_f,
\end{equation}
\noindent where $\mathcal{T}$ denotes the Richards–Wolf inverse mapping operator, which effects the projection from the spherical polarization basis of the scattered field onto the Cartesian polarization basis of the collimated LO, and the integration is carried out over the collection solid angle $\Omega_{\mathrm{cl}}$. The spatial mask $\mathcal{M}_q(\mathbf{k}_f)$ mathematically represents the physical layout of the balanced detector mapped into the momentum space of the collection aperture. For transverse position readout, the signal is extracted differentially by subtracting the optical power between opposite halves of the beam. This operation corresponds to an anti-symmetric sign-flip mask, specifically
\begin{equation}
    \mathcal{M}_{x/y}(\mathbf{k}_f) = \mathrm{sgn}(k_{f,x/y}),
\end{equation}
\noindent which selectively isolates the transverse phase gradients while rejecting symmetric intensity noise. Conversely, longitudinal motion along the optical axis induces a symmetric phase shift across the wavefront; extracting this signal requires integrating the total interference over the entire aperture, corresponding to a uniform scalar mask $\mathcal{M}_z(\mathbf{k}_f) = 1$.

The hardware detection efficiency \(\eta_{\mathrm{det},q}\) is subsequently defined as the signal-to-noise ratio (SNR) associated with this measurement, normalized by both the absolute geometric information bound \(C_q\) and the total reference (LO) power:
\begin{equation}
    \eta_{\mathrm{det},q} 
    = \frac{1}{C_q}\,
      \frac{ I_q^2 }
           { \displaystyle \int_{\Omega_{\mathrm{cl}}} 
               \lvert \mathbf{E}_{\mathrm{LO}} \rvert^2 \, d\Omega_f }.
\end{equation}

Evaluating this metric with the ASR model reveals non-paraxial effects that drive the detection efficiency away from the PWA analytical predictions. Figure~\ref{fig:measurementEfficiency} quantifies these deviations by comparing the ASR results with analytical bounds obtained from the PWA~\cite{tebbenjohanns2019optimal} for both forward and backward detection geometries. For the backward detection calculations presented here, we assume the local oscillator is identical to the trapping field in both spatial distribution and polarization; however, the generalized formalism allows for the selection of a completely arbitrary local oscillator. The analysis identifies two qualitatively distinct regimes as a function of the objective filling factor $f_0$.

In the limit of a strongly overfilled aperture ($f_0 = 10.0$), the incident Gaussian beam approaches a uniform top-hat intensity distribution, closely reproducing the plane--wave-like illumination assumed in the analytical PWA. In this regime, the ASR-derived detection efficiencies associated with the transverse axis orthogonal to the trapping polarization ($\eta_y$) and with the longitudinal axis ($\eta_z$) converge exactly to the corresponding PWA bounds (Fig.~\ref{fig:measurementEfficiency}c,d). In contrast, the efficiency along the measurement axis aligned with the trapping polarization ($\eta_x$) exhibits a pronounced divergence from the scalar theory due to focal depolarization. At high numerical aperture, strong focusing inevitably transforms a substantial fraction of the incident $x$-polarized field into a longitudinal $z$-polarized component at the focus, causing the nanoparticle to radiate partially as a $z$-oriented dipole. In the forward detection scheme (Fig.~\ref{fig:measurementEfficiency}c), this depolarization manifests as a measurement penalty; because the forward collection lens cannot effectively map this $z$-dipole radiation into the transverse local oscillator mode, the signal is lost. Consequently, the scalar PWA overestimates the forward $\eta_x$ by implicitly assuming all scattered photons remain available for interference. Conversely, in the backward detection scheme (Fig.~\ref{fig:measurementEfficiency}d), the exact ASR efficiency exceeds the PWA bound. The trapping objective, acting in reverse to collimate the back-scattered light, geometrically rotates the high-angle $z$-dipole radiation back into the transverse plane. This coherently recovered light constructively interferes with the $x$-polarized local oscillator. Because the scalar PWA completely ignores both the generation of the longitudinal field and its subsequent coherent recovery by the high-NA collection optics, it underestimates the backward detection efficiency.

When a realistic Gaussian input mode is used ($f_0 = 1.0$), the analytical PWA overestimates the detection efficiency along all three spatial directions (Fig.~\ref{fig:measurementEfficiency}a,b). This systematic discrepancy originates from the non-uniform angular spectrum of a Gaussian beam, which suppresses the power carried by high-angle marginal rays relative to a top-hat profile at the same NA. The resulting angular apodization increases the focal spot size and attenuates the steep spatial field gradients that encode positional information. As a result, the actual spatial information content of the scattered field is reduced compared to that predicted by the idealized uniform-illumination PWA, leading to a global decrease in the achievable mode-matching efficiency.

\subsection{Total measurement efficiency}

While the mode-matching efficiency $\eta_{\mathrm{det},q}$ is fundamentally determined by the focal geometry and the vectorial structure of the scattered electromagnetic field, practical measurements are further limited by downstream optical losses, collectively denoted by $\eta_{\mathrm{opt},q}$. This lumped parameter encapsulates the cumulative signal attenuation introduced by the experimental apparatus, including scattering and absorption in all subsequent optical components, as well as the finite quantum efficiency of the balanced photodiodes. These imperfections lead to an additional reduction of the coherent signal, such that the effective measurement rate is given by
\begin{equation}
    \Gamma_{\mathrm{M},q} = \eta_{\mathrm{opt},q}\,\eta_{\mathrm{det},q}\,\Gamma_{\mathrm{S}}
\end{equation}

As a consequence, the effective imprecision noise is enhanced to $S_{I,q}^{\mathrm{eff}} = S_{I,q}/(\eta_{\mathrm{opt},q}\eta_{\mathrm{det},q})$. Ultimately, these combined hardware limitations cause the performance of a realistic experimental system to deviate from the ideal Heisenberg limit by exactly the total efficiency factor:
\begin{equation}
    S_{I,q}^{\mathrm{eff} S_{\mathrm{bk},q}} = \frac{1}{\eta_{\mathrm{opt},q}\eta_{\mathrm{det},q}}\,\frac{\hbar^2}{4}.
\end{equation}

\section{Conclusions}\label{sec:conclusions}

In this work, we establish a generalized vectorial theoretical framework to quantify the fundamental limits imposed by measurement backaction and detection efficiency in levitated optomechanical systems. Specifically, we develop a scattering theory valid in the dipole regime, based on a semiclassical description in which the trapping field is modeled as a flux of discrete photons obeying Poissonian statistics. Within this model, we evaluate the momentum exchange as well as the associated phase shifts of the scattered field, which encode and transport information about the particle’s position. We show that a single parameter, the generalized geometric factor $\mathrm{C}_q$, encapsulates the complete spatial and vectorial structure of the problem and thereby fundamentally determines both the backaction and the measurement imprecision noise.

Building upon this fundamental limit, we investigated how practical experimental configurations depart from the absolute Heisenberg bound. In particular, we modeled the hardware detection efficiency $\eta_{\mathrm{det},q}$ by employing the Richards–Wolf inverse mapping to project the spherical far-field dipole radiation onto the Cartesian basis of a collimated local oscillator. This fully vectorial approach allowed us to accurately simulate spatial mode-matching for both forward and backward collection geometries, demonstrating that scalar approximations systematically overestimate achievable measurement efficiencies.

To illustrate the physical insights enabled by this vectorial formalism, we analyzed the performance of a radially polarized trapping beam. Unlike a standard linearly polarized Gaussian tweezer, which suffers from focal depolarization, the high-NA focusing of a radial mode yields a purely longitudinal, $z$-polarized local field at the trap center. The nanoparticle thus acts as a $z$-oriented dipole, radiating its energy predominantly into the transverse plane. Because these transversely scattered photons carry negligible longitudinal momentum, the random momentum recoil imparted to the particle along the optical axis is severely restricted. This geometric orthogonality between the dipole emission pattern and the optical axis fundamentally suppresses the axial measurement backaction. Concurrently, the structured momentum gradients of the radial field reshape the information radiation pattern, demonstrating that complex vectorial beams can be tailored to route positional information into more advantageous solid angles.

Finally, to support the levitated optomechanics community in designing near-Heisenberg-limited experiments and exploring novel vector-beam configurations, the theoretical and numerical methods developed in this work are provided as an open-source Python package, the LevitationToolbox~\cite{tandeitnik2026levitationtoolbox}.

\section*{Acknowledgments}

We acknowledge support from the Coordenac\~ao de Aperfei\c{c}oamento de Pessoal de N\'ivel Superior - Brasil (CAPES) - Finance Code 001, the Brazilian National Institute of Science and Technology in Quantum Devices (INCT-DQ) and the Brazilian National Council for Scientific and Technological Development (CNPq, Grant No. 408783/2024-9), Funda\c{c}\~ao de Amparo \`a Pesquisa do Estado do Rio de Janeiro (FAPERJ Scholarships No. E-26/203.727/2025, E-26/204.170/2024,  E-26/210.077/2023, E-26/200.251/2023, E-26/210.249/2024, E-26/210.824/2025 and E-26/210.373/2026), Funda\c{c}\~ao de Amparo \`a Pesquisa do Estado de São Paulo (FAPESP processo 2021/06736-5), the Serrapilheira Institute (grant No. Serra – 2211-42299) and StoneLab.


\bibliography{main}

\newpage

\appendix
\onecolumngrid 

\end{document}